\documentclass[fleqn]{llncs}
\usepackage{lmodern}
\usepackage[T1]{fontenc}
\pdfoutput=1
\usepackage{graphicx}
\usepackage{pslatex}
\usepackage{pst-node}
\usepackage{verbatim}
\usepackage{color}
\usepackage{amsmath}
\usepackage{amsfonts}
\usepackage{subfigure}

\usepackage{algorithm}
\usepackage[noend]{algorithmic}

\newcommand{\keywords}[1]
{\par\addvspace\baselineskip\noindent\keywordname\enspace\ignorespaces#1}

\title{FUSE: Multiple Network Alignment via Data Fusion}
\author{ Vladimir Gligorijevi\'{c}\thanks{Both authors contributed equally}
\and No\"el Malod-Dognin$^{\star}$ \and Nata\v{s}a Pr\v{z}ulj\thanks{Corresponding author.}}
\institute{ Department of Computing, Imperial College London, United Kingdom.\\
\tt{natasha@imperial.ac.uk}}

\begin{document}

\maketitle

\begin{abstract}
Discovering patterns in networks of protein-protein interactions
(PPIs) is a central problem in systems biology. Alignments between
these networks aid functional understanding as they uncover important
information, such as evolutionary conserved pathways, protein
complexes and functional orthologs. However, the complexity of the
multiple network alignment problem grows exponentially with the number
of networks being aligned and designing a multiple network aligner
that is both scalable and that produces biologically meaningful
alignments is a challenging task that has not been fully
addressed. The objective of a multiple network alignment is to create
clusters of nodes that are evolutionarily conserved and functionally
consistent across all networks. Unfortunately, the alignment methods
proposed thus far do not fully meet this objective, as they are guided
by pairwise scores that do not utilize the entire functional and
topological information across all networks.

To overcome this weakness, we propose FUSE, a multiple network aligner
that utilizes all functional and topological information in all PPI
networks.  It works in two steps.  First, it computes novel similarity
scores of proteins across the PPI networks by fusing from
\textit{all} aligned networks both the protein wiring patterns and
their sequence similarities.  It does this by using Non-negative
Matrix Tri-Factorization (NMTF). When we apply NMTF on the five
largest and most complete PPI networks from BioGRID, we show that NMTF
finds a larger number of protein pairs across the PPI networks that 
are functionally conserved than can be found by using protein sequence 
similarities alone.  This demonstrates complementarity of protein 
sequence and their wiring patterns in the PPI networks. In the second 
step, FUSE uses a novel maximum weight $k$-partite matching approximation 
algorithm to find an alignment between multiple networks. We compare FUSE 
with the state of the art multiple network aligners and show that it 
produces the largest number of functionally consistent clusters that cover 
all aligned PPI networks. Also, FUSE is more computationally efficient than 
other multiple network aligners.

\end{abstract}

\keywords{Multiple network alignment, $k$-partite matching, data fusion, non-negative matrix tri-factorisation}

\section{Introduction}
Understanding the patterns in molecular interaction networks is of foremost importance in systems biology, as it is 
instrumental to understanding the functioning of the cell \cite{ryan13}. A large number of studies focused on 
understanding the topology of these networks \cite{przulj11,mitra13}.
Network alignment started as a pairwise problem: given two networks, aligning them means finding a node-to-node 
mapping (called an {\em alignment}) between the networks that groups together evolutionarily or functionally 
related proteins between the networks. These methods uncovered valuable information, such as evolutionarily 
conserved pathways and protein complexes \cite{kelley03,kuchaiev10}, and functional orthologs \cite{bandyopadhyay06}.
Finding these allows transfer of information across species, such as performing Herpes viral experiments in yeast 
or fly and then applying the insights towards understanding the mechanisms of human diseases \cite{uetz06}.

The pairwise network alignment problem is computationally intractable due to NP-completeness of the underlying sub-graph 
isomorphism problem \cite{cook71}. Hence, several pairwise network alignment heuristics have been proposed. Early methods, 
called {\em local network aligners}, search for small, but highly conserved sub-networks \cite{kelley04,koyuturk06,flannick06}. 
As such sub-networks can be duplicated, local network aligners often produce one-to-many or many-to-many mappings, in 
which a node from a given network can be mapped to several nodes of the other network. While these multiple mappings can 
indicate gene duplications, they are often biologically implausible \cite{singh07}. Hence, {\em global network aligners}, 
which perform an overall comparison of the input networks and produce one-to-one mappings between the nodes of the two 
networks have been introduced (see \cite{clark14} for the most recent comparison of pairwise network aligners).

The number of known protein-protein interactions (PPIs) increased dramatically over the last two decades thanks to the 
technological advances in high-throughput interaction detection techniques, such as yeast two-hybrid \cite{ito00,uetz00} 
and affinity purification coupled to mass spectrometry \cite{ho02}. With the availability of PPI networks of multiple 
species came the multiple network alignment problem, where given $k$ networks, aligning them means to group together the 
proteins that are evolutionarily or functionally conserved between the networks. Similar to pairwise network alignment, 
multiple network alignment can be local or global, with node mappings one-to-one or many-to-many. As the complexity of 
the problem grows exponentially with the number of networks to be aligned, the proposed multiple network alignment 
algorithms use simple and scalable alignment schemes. The pioneering multiple network alignment algorithm is NetworkBLAST 
\cite{sharan05,kalaev08}, which greedily searches for highly conserved local regions in the alignment graph constructed 
from the pairwise protein sequence similarities. Graemlin \cite{flannick06} produces local multiple network alignments 
using a progressive alignment scheme, by successively performing pairwise alignments of the closest network pairs. IsoRank 
\cite{singh08} and its successor IsoRankN \cite{liao09} are the first multiple network aligners that do not only use 
pairwise sequence similarity to guide their alignment processes, but also take into account the topology (i.e., wiring 
patterns) around the two nodes in their corresponding networks to build up global many-to-many multiple network alignments, 
using a derivative of Google's PageRank algorithm. Smetana \cite{sahraeian13} also produces global many-to-many multiple 
network alignments using both pairwise sequence scores and pairwise topological scores, which are derived from a 
semi-Markov random walk model. While NetCoffee \cite{hu13} does not use topological information to build its global 
one-to-one alignment, it is the first multiple network aligner in which the score for mapping two nodes does not only 
depend on the scores in pairs of networks, but also on their conservation across all PPI networks being aligned, by using 
a triplet approach similar to the multiple sequence aligner, T-Coffee \cite{notredame00}. Finally, Beams \cite{alkan14} 
is a fast heuristics that constructs global many-to-many multiple network alignments from the pairwise sequence
similarities of the nodes by using a backbone (seed) extraction and merge strategy. In the above mentioned aligners, most 
of the node mapping scores are local, in the sense that they only consider the sequence similarity or the topological 
similarity of the nodes. The only exception is NetCoffee, but its global scores are only based on sequence similarity 
and do not take into account the topology of the networks.

To overcome these limitations, we propose FUSE, a novel multiple
network aligner that consists of two parts.  In the first part, we
compute novel similarity (association) scores between proteins by
fusing sequence similarities and network wiring patterns over {\em
all} proteins in {\em all} PPI networks being aligned.  We do this by
using Non-negative Matrix Tri-Factorization (NMTF) technique
\cite{wang11}, initially used for co-clustering heterogeneous data,
but recently proposed as a data fusion technique as well. NMTF has
demonstrated a great potential in addressing various biological
problems, such as drug-induced liver injuries prediction
\cite{zitnik14a}, disease association prediction \cite{zitnik13} and
gene function prediction \cite{gligorijevic14,zitnik14b}.  We apply
NMTF on the PPI networks of the five species that have the largest and
the most complete sets of PPIs in BioGRID database \cite{chatr13}.  On
this dataset, the fusion process changes the values of sequence
similarities between proteins based on network topologies, so that
some of the sequence similarities that existed before fusion disappear
(about 41\% in our experiments), while a large set of new ones is
created by the \textit{matrix completion} property of NMTF
\cite{koren09}. We show that the set of protein pairs predicted to be
similar by NMTF, which contains 38 times more pairs than the set of
sequence-similar pairs due to fusion with network topology, has the
same functional consistency (i.e., shared GO terms across the pairs)
as the set of protein pairs found to be similar by sequence
alignment. To avoid losing sequence similarity information, our final
\emph{functional similarity score} for a pair of protein is a weighted
sum of the sequence similarity and the similarity predicted by
NMTF. 

In the second part of FUSE, to construct a global one-to-one multiple
network alignment, first we construct an edge-weighted $k$-partite
graph, with the proteins of each of the $k$ PPI networks being
partitions of its node set and the above described functional
similarity scores being edge weights.  To construct a multiple network
alignment, we find a maximum weight $k$-partite matching in this
graph. As finding a maximum weight $k$-partite matching is NP-hard
\cite{karp72}, we propose a novel approximation algorithm for it. 

We evaluate the performance of FUSE against other state of the art
multiple network aligners and show that FUSE produces the largest
number of functionally consistent clusters that map proteins over all
aligned networks.  Moreover, we show that FUSE is scalable and
computationally more efficient than all of the previous aligners
except Smetana (but Smetana's aligned proteins are not as functionally
consistent as FUSE's; detailed below). Specifically, the data-fusion
step is the most time consuming in FUSE with the time complexity of
$O(\nu^3)$, where $\nu$ is the number of proteins in the largest PPI 
network being aligned, while the alignment step has a smaller time complexity
of $O(kn^2\log{n} + kne)$, where $n$ is the number of proteins in all PPI networks and $e$ is the number of functional
associations (similarity scores) between them.



\section{Materials and methods}

\subsection{Datasets}
From BioGRID (v3.2.111, April 25th, 2014) \cite{chatr13},
we obtained the PPI networks of the 5 organisms having the largest and the most complete sets of physical PPIs: 
 {\em Homo sapiens} (HS), {\em Saccharomyces cerevisiae} (SC), {\em Drosophila melanogaster} (DM), {\em Mus musculus} (MM), and {\em Caenorhabditis elegans} (CE).
We retrieved the corresponding protein sequences from NCBI's Entrez Gene database \cite{maglott05}
and computed their pairwise similarities using BLAST \cite{altschul90}.
We also retrieved from NCBI's Entrez Gene database the Gene Ontology (GO) annotations of the proteins.
Note that we only used experimentally validated GO annotations (i.e, excluding the annotations from computational analysis evidence such as sequence similarity)
and that we additionally excluded annotations derived from protein-protein interaction experiments (code IPI).
To standardize the GO annotations of proteins, similar to the evaluation methods of \cite{singh08,liao09,alkan14},
we restrict the protein annotations to the fifth level of the GO directed acyclic graph by ignoring the higher-level annotations and replacing the deeper-level
annotations with their ancestors at the restricted level.
The network statistics are detailed in Table \ref{table:PPIs}.

\begin{table}
	\begin{center}
		\begin{tabular}{| l || r | r | r | r || r |}
			\hline
			Id	& \# Nodes	& BP Ann. (\%)	& MF Ann. (\%)	& CC Ann. (\%)	& \# Edges \\
			\hline
			HS	& 14,164	& 37.2		& 23.2		& 9.6		& 127,907\\
			SC	& 6,004		& 65.0		& 41.7		& 17.4		& 223,008\\
			DM	& 8,125		& 36.1		& 13.4		& 6.3		& 38,892\\
			MM	& 5,105		& 53.3		& 23.9		& 10.6		& 11,061\\
			CE	& 3,841 	& 35.0		& 7.3		& 4.2		& 7,726\\
			\hline
		\end{tabular}
		\vspace{0.2cm}\caption{{\bf The five PPI networks considered in this study.}
		For each PPI network (row), the table presents its Id (column 1), its number of nodes (column 2),
		its percentage of nodes that are annotated with at least one GO term from either biological process category (BP, column 3),
		molecular function category (MF, column 4), or cellular component (CC, column 5), and finally, its number of edges (column 6).}\label{table:PPIs}
		\vspace{-0.5cm}
	\end{center}
\end{table}

\subsection{Method}
The PPI of each species $i$ is represented by a graph (network), $N_i = (V_i, E_i)$, where 
the nodes in $V_i$ represent proteins, and where two proteins are connected by an edge in 
$E_i$ if they interact. Our multiple network alignment strategy consists of two steps. In 
the next two paragraphs, we give a short overview of these steps, before giving the full 
details of the methodology.

First, we use all PPI networks to be aligned and all the protein sequence similarities between 
them, as inputs into the NMTF-based data fusion technique to compute new protein {\em  
similarity} scores between the proteins of the networks. Considering the obtained normal 
distribution of similarity scores for aligning the 5 PPI networks described above, we define 
as {\em significant} the scores that are in top 5\%. We combine significant scores with the 
original sequence similarities to derive the final {\em functional scores} between pairs of 
proteins for the reasons explained is section \ref{sec:res_nmtf}. We construct an edge-weighted 
$k$-partite graph $G=( \bigcup^{k}_{i=1} V_i, E, W)$, where the node set is the union of the 
nodes sets (proteins) $V_i$ of the input PPI networks; two nodes $u\in V_i$, $v \in V_j$, 
$i\neq j$, are connected by an edge $(u,v)$ in $E$ if their functional score is greater than 
zero; the corresponding edge weight in $W$ is their functional score. No edge exists between 
nodes coming from the same subset $V_i$ by definition of a $k$-partite graph.

Second, we construct a one-to-one global multiple network alignment by using an approximate 
maximum weight $k$-partite matching solver on $G$. 

\subsubsection{Non-negative matrix tri-factorization.}
NMTF is a machine learning technique initially designed for co-clustering of 
multi-type relational data \cite{wangli08,wang11}. In this paper, we consider 
proteins belonging to different species as different data types. In the case of two species, 
$i$ and $j$, the sequence similarity scores between their proteins are recorded 
in the high-dimensional relation matrix, $\mathbf{R}_{ij} \in \mathbb{R}^{n_i \times n_j}$, 
where, $n_i$ is the number of proteins in the species $i$ and $n_j$ is the number of proteins 
in the species $j$. Entries in the relation matrix are $e$-values of the protein sequence 
alignments computed by using BLAST. Specifically, we use $-log(eval)$ as a measure of association 
between protein pairs. NMTF estimates the high-dimensional matrix, $\mathbf{R}_{ij}$ as a product of 
low-dimensional non-negative matrix factors: 
$\displaystyle \mathbf{R}_{ij} \approx \mathbf{G}_i\mathbf{S}_{ij}\mathbf{G}_j^{T}$, 
where, $\mathbf{G}_i \in \mathbb{R}_{+}^{n_i \times k_i}$ and 
$\mathbf{G}_j \in \mathbb{R}_{+}^{n_j \times k_j}$ correspond to the cluster indicator 
matrices of proteins in the first and the second species respectively, and 
$\mathbf{S}_{ij} \in \mathbb{R}^{k_i \times k_j}$ is a low-dimensional, compressed 
version of $\mathbf{R}_{ij}$, where the choice of rank parameters $k_i, k_j \ll min\{n_1,n_2\}$ 
provides dimensionality reduction. The close connection between non-negative matrix factorization 
problem and the clustering problem is well established \cite{wang11,ding06,ding05}. 

In addition to co-clustering, NMTF technique can also be used for \textit{matrix completion}. 
Namely, some entries in the initial relation matrix $\mathbf{R}_{ij}$ are zero (due to lack of 
sequence similarities between the corresponding proteins) and they can be recovered from the obtained 
low-dimensional matrix factors using the \textit{reconstructed relation matrix}: $\mathbf{\hat{R}}_{ij} 
= \mathbf{G}_i\mathbf{S}_{ij}\mathbf{G}_j$ (detailed bellow). Here we use this property to predict 
new and recover the existing association between proteins. To obtain the low-dimensional matrix factors, 
$\mathbf{G}_i, \mathbf{S}_{ij}, \mathbf{G}_j$, we solve the following optimization problem:
\begin{equation*}
\min_{\mathbf{G}_i \geq 0, \mathbf{G}_j \geq 0} J =  \parallel \mathbf{R}_{ij} - \mathbf{G}_i \mathbf{S}_{ij} \mathbf{G}_j^{T} \parallel^{2}_{F}
\end{equation*}
    
We incorporate PPI network topology as constraints into our optimization problem; violation of these 
constraints causes penalties to our objective function. This is motivated by the co-clustering 
problem which uses networks as prior information to cluster proteins. Namely, the aim is to allow 
proteins interacting within a PPI network to belong to the same cluster. Interactions between 
proteins in PPI network, $i$, are represented by a graph Laplacian matrix, 
$\mathbf{L}_i = \mathbf{D}_i - \mathbf{A}_i$, where $\mathbf{A}_i$ is the adjacency matrix of network 
$i$ and $\mathbf{D}_i$ is the diagonal degree matrix of $i$ (i.e., diagonal entries in $\mathbf{D}_i$ 
are row sums of $\mathbf{A}_i$). For all five of our PPI networks we construct a Laplacian 
matrix, resulting in the set: $\{\mathbf{L}_1,\dots,\mathbf{L}_5\}$. 

We use a block-based representation of relation ($\mathbf{R}$) and Laplacian ($\mathbf{L}$) matrices 
and matrix factors ($\mathbf{S}$ and $\mathbf{G}$) for our 5 PPI networks as follows:
{   \scriptsize \begin{align*}
    \label{block}
    \mathbf{R} = \begin{bmatrix}
    0 & \mathbf{R}_{12} & \ldots & \mathbf{R}_{15}\\
    \mathbf{R}^{T}_{12} & 0 & \ldots & \mathbf{R}_{25}\\
    \vdots & \vdots &  \ddots & \vdots\\\
    \mathbf{R}^{T}_{15} & \mathbf{R}^{T}_{25} & \ldots & 0
    \end{bmatrix} \text{, } \quad & 
    \mathbf{L} = \begin{bmatrix}
    \mathbf{L}_{1} & 0 & \ldots & 0\\
    0 & \mathbf{L}_{2} & \ldots & 0\\
    \vdots & \vdots & \ddots & \vdots\\ 
    0 & 0 &  \ldots & \mathbf{L}_{5}  
    \end{bmatrix} \text{; } \quad &
    \mathbf{S} = \begin{bmatrix}
    0 & \mathbf{S}_{12} & \ldots & \mathbf{S}_{15}\\
    \mathbf{S}^{T}_{12} & 0 & \ldots & \mathbf{S}_{25}\\
    \vdots & \vdots &  \ddots & \vdots\\\
    \mathbf{S}^{T}_{15} & \mathbf{S}^{T}_{25} & \ldots & 0
    \end{bmatrix}  \text{, } \quad &
    \mathbf{G} = \begin{bmatrix}
    \mathbf{G}_{1} & 0 & \ldots & 0\\
    0 & \mathbf{G}_{2} & \ldots & 0\\
    \vdots & \vdots & \ddots & \vdots\\ 
    0 & 0 &  \ldots & \mathbf{G}_{5}
    \end{bmatrix} 
\end{align*}}	
\noindent To simultaneously factorize all relation matrices, 
$\displaystyle \mathbf{R}_{ij} \approx \mathbf{G}_i\mathbf{S}_{ij}\mathbf{G}_j^{T}$, 
$0 \leq i,j \leq 5$, under the constraints of PPI networks, we minimize 
the following objective function:
\begin{equation}
    \label{objective}
    \min\limits_{\mathbf{G} \geq 0} J   = \big[ \parallel \mathbf{R} - 
    \mathbf{G} \mathbf{S} \mathbf{G}^{T}
    \parallel^{2}_{F} + \gamma Tr(\mathbf{G}^{T} \mathbf{L} \mathbf{G}) \big]
\end{equation}
\noindent where $Tr$ denotes the trace of a matrix and $\gamma$ is a regularization 
parameter which balances the influence of network topologies in reconstruction of 
the relation matrix. The second term of equation \ref{objective} is the penalization term. 
It takes into account protein connections within the PPI network in the following way: connected 
pairs of proteins are represented with negative entries in the Laplacian matrix of the 
corresponding PPI network, and these entries act as rewards that reduce the value of the 
objective function, $J$, forcing the proteins to belong to the same cluster.

The optimization problem (Equation \ref{objective}) is solved by applying the algorithm  
following \textit{multiplicative update rules} used to compute matrices $\mathbf{G}$ and 
$\mathbf{S}$ and under which the objective function, $J$, is non-increasing \cite{wangli08}. 
These update rules are derived by minimizing the \textit{Langragian function}, $\mathcal{L}$, 
constructed from the objective function and all additional constraints, including positivity 
of matrix factors $\mathbf{G}$, as follows \cite{wangli08}. The update rule for $\mathbf{S}$ 
is obtained by fixing the other matrix factor, $\mathbf{G}$, and finding the roots of the 
equation: $\displaystyle {\partial \mathcal{L}} / {\partial \mathbf{S} = 0}$. A similar 
procedure is followed for obtaining the update rule for matrix factor $\mathbf{G}$. The 
multiplicative update rules, their derivation and the proof of their convergence can be 
found in \cite{wangli08}.

The central idea of the NMTF-based data fusion approach lies in the fact that the relation 
matrices are not factorized separately, but instead, are coupled by the same 
low-dimensional matrix factors, $\mathbf{G}_i$, which participate in their simultaneous 
decomposition \cite{zitnik13} (see left panel of Figure \ref{fig:fig1} for an illustration). This 
corresponds to the \textit{intermediate} data fusion approach (which keeps the structure 
of the data while inferring a model), that has been shown to be the most accurate from all 
data fusion approaches \cite{lanckriet04,gevaert06,zitnik13}.  

In our study, we use the the following values of parameters for NMTF: (a) factorization ranks, 
$k_1 = 80  $, $k_2 = 90 $, $k_3 = 80 $, $k_4 = 70 $ and $k_5 = 50 $, which we estimated by 
computing principal components of relation matrices by using Principal Component Analysis (PCA) 
\cite{jolliffe05}; (b) we chose regularization parameter, $\gamma = 0.01$, since it gave the 
best biological quality of predicted associations (we tested NMTF for $\gamma \in \{0.001, 0.01, 
0.5, 1.0 \}$).  

After convergence of NMTF, we compute reconstructed relation matrices: $\mathbf{\hat{R}}_{ij} = 
\mathbf{G}_i\mathbf{S}_{ij}\mathbf{G}_j$ for each pair of networks, $i$ and $j$. We compute the 
functional scores of associations between proteins from the statistically significant scores 
($p < 0.05$) predicted from the reconstructed matrices to which we add sequence
similarity scores. 
We do this to avoid losing sequence information since a large number of initial associations is not 
recovered after NMTF procedure (see Section \ref{sec:res_nmtf} for details).  


\subsubsection{Approximate maximum weight $k$-partite matching.}
Using the weighted k-partite graph representation described above,
we globally align multiple networks by finding a maximum weight $k$-partite matching in $G$ (defined above).
Maximum weight $k$-partite matching is known to  be NP-hard for $k \geq 3$ \cite{karp72,papadimitriou94}.
Hence, we approximate it as follows.

We modify the algorithm proposed by He et al. \cite{he00}
for finding an approximate solution to the $k$-partite matching problem.
We define the following graph merge operation.
Let $G=(\bigcup^{k}_{i=1} V_i, E, W)$ be an edge-weighted $k$-partite graph, and $G[V_i, V_j]$
be the edge-weighted bi-partite subraph of $G$ that is induced by the two subsets of nodes $V_i$ and $V_j$.
Let $F_{i,j}$ = $\{u_1 \leftrightarrow v_1, u_2 \leftrightarrow v_2, \dots, u_l \leftrightarrow v_l\}$ be a matching of $G[V_i, V_j]$,
where $u_k \leftrightarrow v_k$ means that node $u_k \in V_i$ is matched with node $v_k \in V_j$.
We merge $V_j$ with $V_i$ into $V_{ij}$ by identifying the mapped nodes $u_k \leftrightarrow v_k$ and
by creating a corresponding {\em merged node} $u_{k}v_{k} \in V_{ij}$.
These merged nodes inherit the edges from their parent nodes, and multiple edges are replaced by a single edge with
the sum of weights of the multiple edges as the new weight of the edge.
We also move into $V_{ij}$ the nodes of $V_i$ and $V_j$ that are not matched.
The new weighted graph $G_{ij}$ is called the {\em merge} of $V_j$ to $V_i$ from $G$ along
$F_{i,j}$. We note that $G_{ij}$ is an edge-weighted $(k-1)$ partite graph.

Our approximated maximum weight $k$-partite matching algorithm can be seen as a progressive aligner which first maps and merges the two first networks,
and then successively adds into the ``merge graph'' the remaining networks (see Algorithm \ref{algo:kpartitematching}).

\begin{algorithm}
\caption{Approximate maximum weight $k$-partite matching.}
\label{algo:kpartitematching}
\begin{algorithmic}\footnotesize
\STATE{ {\bf Input} $G=(\bigcup^{k}_{i=1} V_i, E, W)$}
\FOR{$i = \{2,\dots, k\}$}
	\STATE{ Find maximum weight bipartite matching $F_{1,i}$ of $G[V_1, V_i]$ }
	\STATE{ Construct $G_{1i}$, the merge of $V_i$ to $V_1$ from $G$ along $F_{1,i}$ }
	\STATE{ Set $G = G_{1i}$, and relabel $V_{1i}$ as $V_1$ }
\ENDFOR
\STATE{$C = \{ \emptyset \}$}
\FOR{ each merged node $u$ in $V_1$}
	\STATE{ Cluster $C_u$ is the set of nodes that are merged into $u$  }
	\STATE{ Add $C_u$ to $C$}
\ENDFOR
\STATE{ {\bf Output} $C$}
\end{algorithmic}
\end{algorithm}

The main operation in Algorithm \ref{algo:kpartitematching} is finding a maximum weight matching in an induced bi-partite graph, which takes $O(n^2 \log{n} + ne)$ time \cite{bondy76,lovasz86},
when the $k$-partite graph has $n$ and $e$ edges. There are $k-1$ such operations, hence Algorithm \ref{algo:kpartitematching} computes an approximate solution
for the maximum weight $k$-partite matching problem in $O(kn^2 \log{n} + kne)$ time.

\section{Results}

\subsection{Biological assessment of NMTF predicted protein similarities}
\label{sec:res_nmtf}
The input data consist of $1,137,508$ sequence similarities between all proteins in the PPI 
networks of the 5 species. Using these similarities as input, the NMTF outputs $38,506,872$ 
significant similarities (top 5\%), obtained from the reconstructed relation
matrices. These significant similarities, resulting from NMTF, cover 
58.61\% of the input sequence similarities (\textit{reconstructed}), while the remaining similarities, 
resulting from NMTF, are \textit{predicted}. 

To estimate the impact of PPI network topology on prediction of protein similarities and 
to understand why 41.39\% of the initial sequence similarities are not reconstructed through 
factorization process, we perform the following experiment: for each \textit{reconstructed}, 
\textit{predicted} and \textit{non-reconstructed} protein pair, we count the number of significant 
sequence similarities between their neighbors in the corresponding PPI networks. For the protein 
pairs with reconstructed sequence similarities, we find that their neighbors share on average $83$ 
significant sequence similarities. In contrast, a much smaller number of sequence similar neighbors, 
$46.5$ on average, is observed for the protein pairs with non-reconstructed similarities. Finally, the 
largest average number of similarities between neighboring proteins, $315.5$, exists for protein pairs 
with predicted similarities. This means that NMTF induces new and reconstructs existing similarities 
between proteins that have many sequence similar neighbors in the corresponding PPI networks. Hence, 
the sequence similarity of protein pairs without many sequence similar neighbors in their PPI networks 
will be lost in NMTF process. 

\begin{figure}[!tpb] 
    \centering
    \begin{tabular}{cc}
	   \includegraphics[width=4cm]{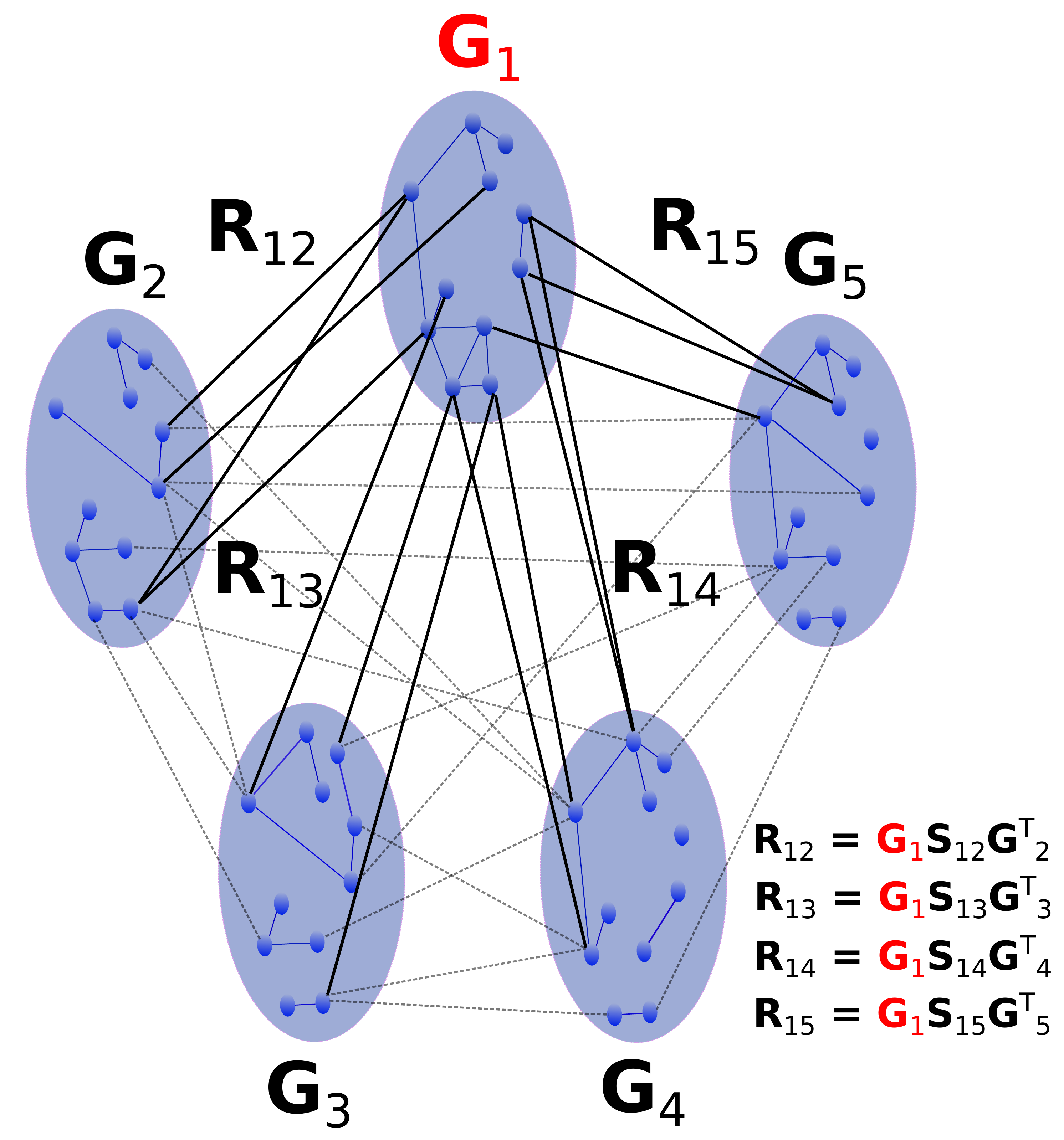}&
       \includegraphics[width=5cm]{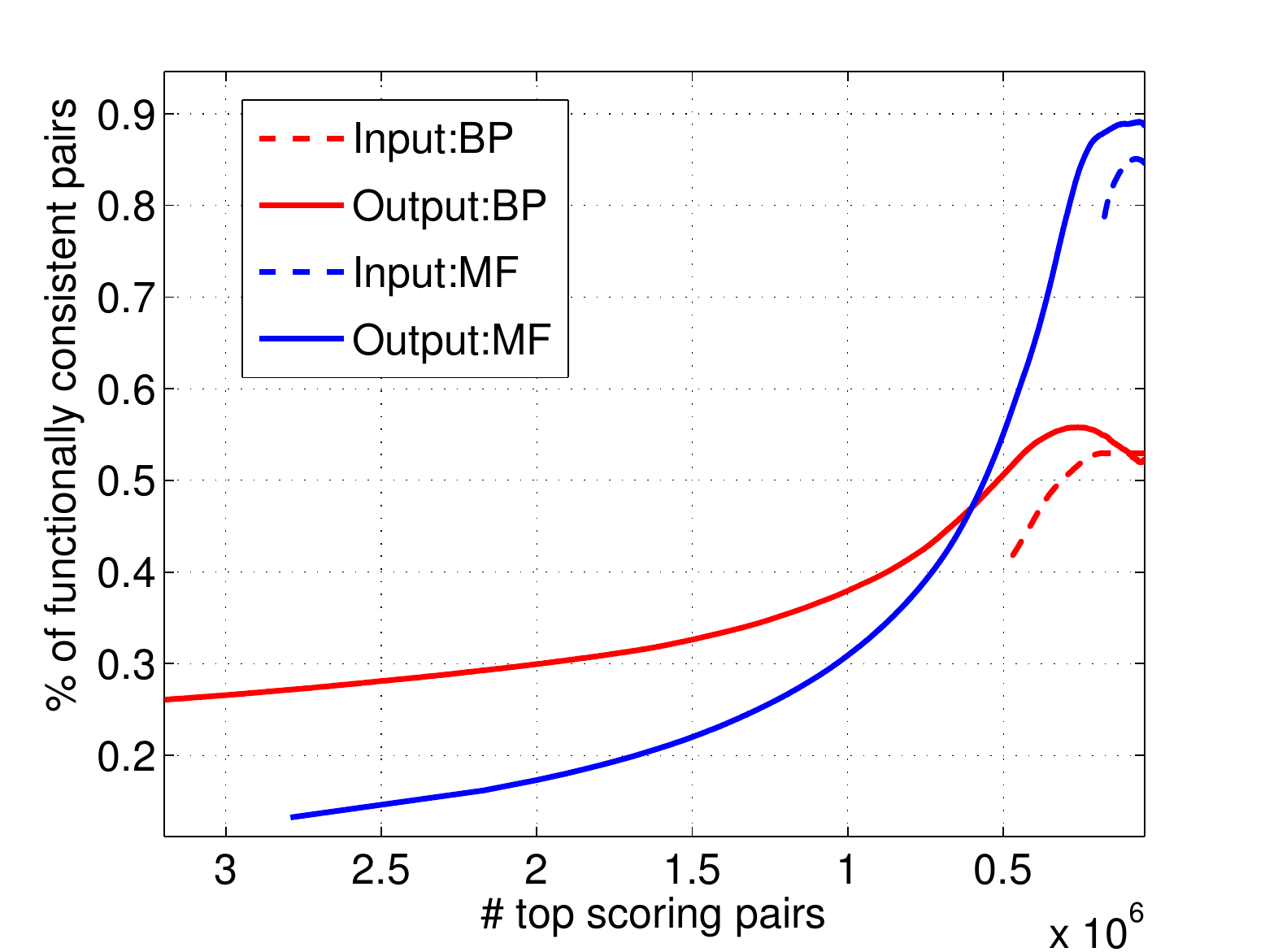}\\
    \end{tabular}
    \caption{Left: An illustration of the basic principle of NMTF-based data
    fusion of 5 PPI networks. Low-dimensional matrix factor $\mathbf{G}_1$, shown in 
    red, is shared in the decompositions of data sets represented by relation matrices: 
    $\mathbf{R}_{12},\mathbf{R}_{13},\mathbf{R}_{14},\mathbf{R}_{15}$. Therefore, the 
    decomposition of $\mathbf{R}_{12}$ depends on the other relation matrices through 
    the shared matrix $\mathbf{G}_1$. Right: GO enrichment assessment of protein similarities 
    predicted by NMTF. Cumulative number of predicted similar protein pairs ($x$-axis) with 
    the percentages of them sharing GO terms ($y$-axis).} 
    \label{fig:fig1}
\end{figure}

To assess the functional consistency of NMTF-predicted protein similarities, we compute the 
the cumulative number of predicted protein pairs and the percentage of them sharing GO term (MF 
and BP  GO terms). Compared with input sequence similar pairs, NMTF produces more functionally 
consistent paired proteins (right panel of Figure \ref{fig:fig1}). This means that topologies of PPI networks 
contribute to functional coherence of protein pairs predicted to be similar by NMTF.

\subsection{FUSE-ing PPI networks}
We FUSE the five PPI networks and assess its results against other multiple network aligners, Beams \cite{alkan14} and Smetana \cite{sahraeian13}.
We tried to obtained alignments from IsorankN \cite{liao09} and NetCoffee \cite{hu13}, but the computations did not finish after more than one week.
We use BLAST e-values as input sequence scores for all methods, using $-log(evalue)$ as the similarity measure.
Both FUSE and Beams use parameter $\alpha \in [0,1]$ to balance the amount of input protein sequence similarity versus network topology.
For these methods, we sample $\alpha$ from 0 to 1, in increments of 0.1.
We left the other parameters of Beams and all the parameters of Smetana at their default values.

\paragraph{\bf Evaluation based on coverage.}

\begin{figure}
	\begin{center}
	\begin{tabular}{c c}
		\multicolumn{2}{c}{ \includegraphics[width=7cm]{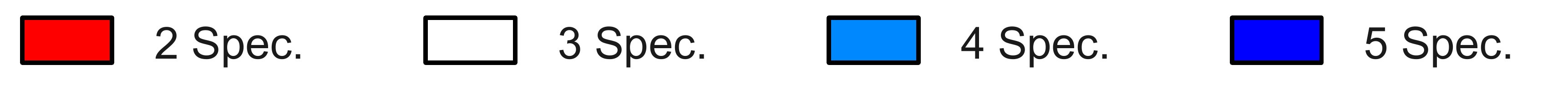}  }\\
		\includegraphics[width=4cm, angle=270]{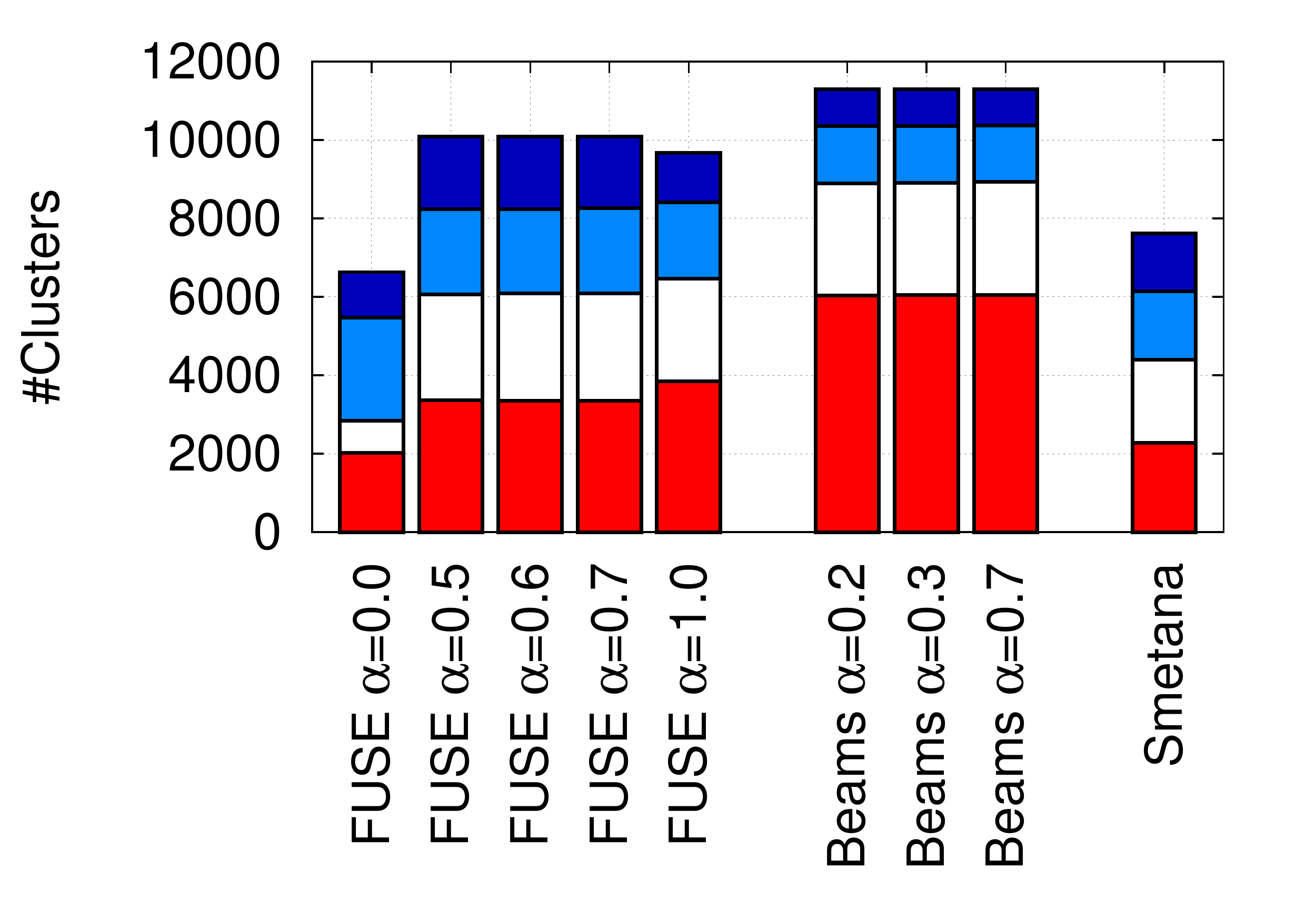} & \hspace{0.5cm}\includegraphics[width=4cm, angle=270]{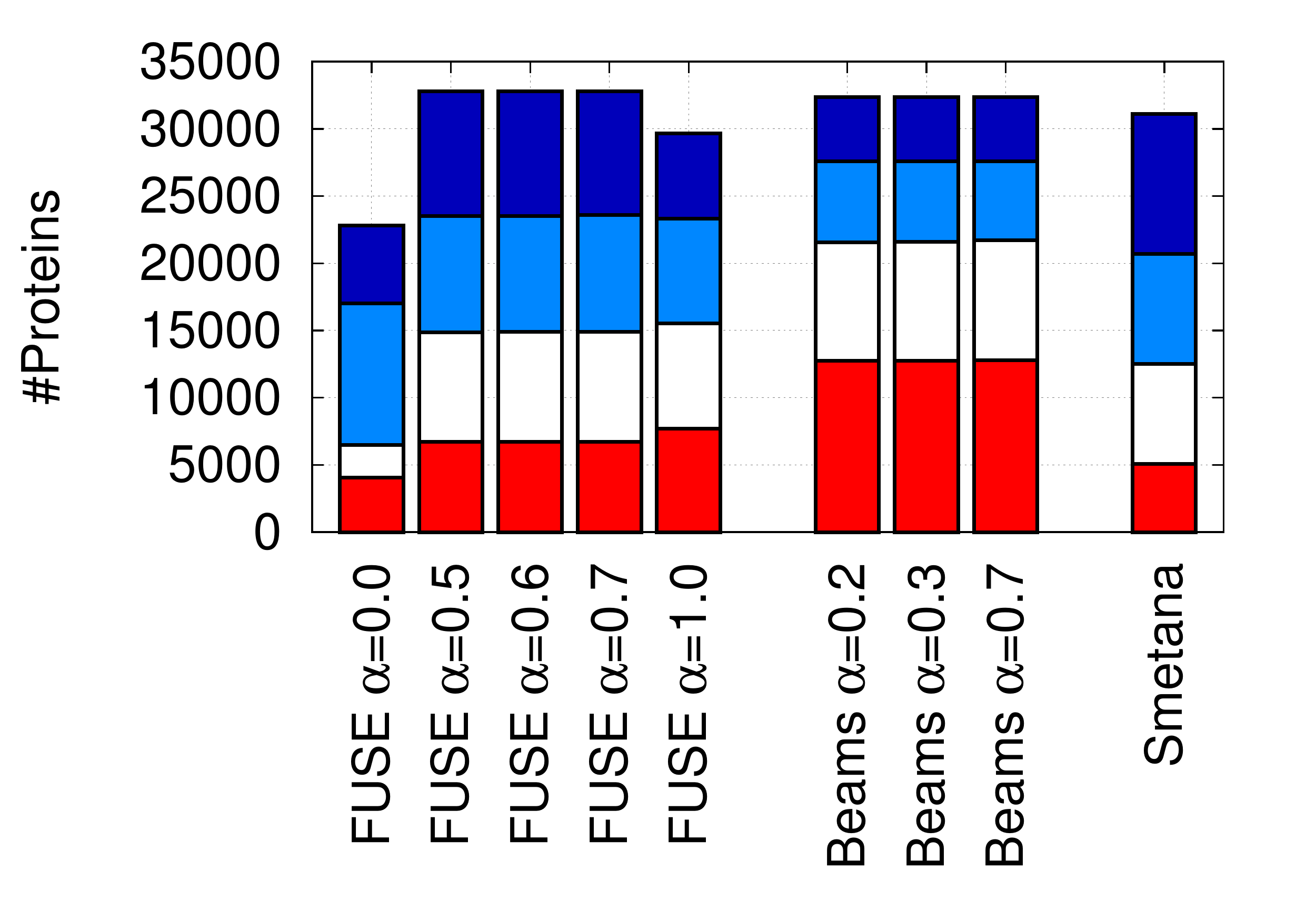} \\
	\end{tabular}
	\caption{{\bf Coverage analysis}. Left: for each alignment produced by the compared aligners (for a specific value of $\alpha$ for FUSE and Beams),
	the bar chart shows the number of clusters containing proteins from $k$ species (see the colour coding on the top). Right: the figure shows the same, but in terms of the number of proteins in these
	clusters.}
	\label{fig:coveragecomp}
	\end{center}
\end{figure}

First, we compare the network aligners on their ability to form protein clusters that cover all of the input PPI networks.
The {\em $k$-coverage} is the number of clusters containing proteins from $k$ different PPI networks. Because the number of proteins per cluster may vary,
the $k$-coverage is also expressed in terms of the number of proteins that are in these clusters.
The {\em total coverage} considers all clusters containing proteins from at least two networks.
The coverage statistics of the alignments are summarised in Figure \ref{fig:coveragecomp}.

FUSE produces larger number of good clusters (containing proteins from all five species, in dark blue in Figure \ref{fig:coveragecomp}), producing 1,855 of such clusters.
Beams achieves the highest total coverage (with up to 11,302 clusters containing proteins from two to five species),
but it does so by producing the largest number of bad clusters (containing proteins from only two species, in red in Figure \ref{fig:coveragecomp}),
producing up to 6,046 of such clusters, and the smallest number of good ones (937 clusters containing 4,803 proteins).

When the coverage is expressed in terms of number of protein in the clusters (right panel of Figure \ref{fig:coveragecomp}), the coverages of FUSE and of Smetana are equivalent.
This means that Smetana puts more proteins in its clusters, and as we show in the next experiment, it does so at the cost of the functional consistency of the clusters.
The total coverage of Beams is equivalent to the one of FUSE and Smetana, but again because Beams puts a larger number of proteins in clusters that covers only two species (with up to 12,805 proteins in these clusters). 

Interestingly, for FUSE, the best coverages are obtained for $\alpha \approx 0.6$, when the functional similarity between proteins is a combination of their
sequence similarity and of their NMTF predicted similarity, which shows the complementarity of network topology and protein sequence as sources of biological information.

\begin{figure}
	\begin{center}
	\begin{tabular}{c c}
		\multicolumn{2}{c}{ \includegraphics[width=7cm]{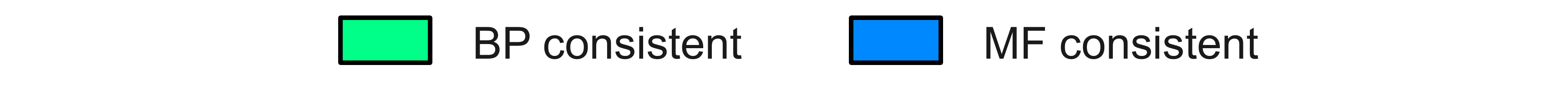}  }\\
		\includegraphics[width=4cm, angle=270]{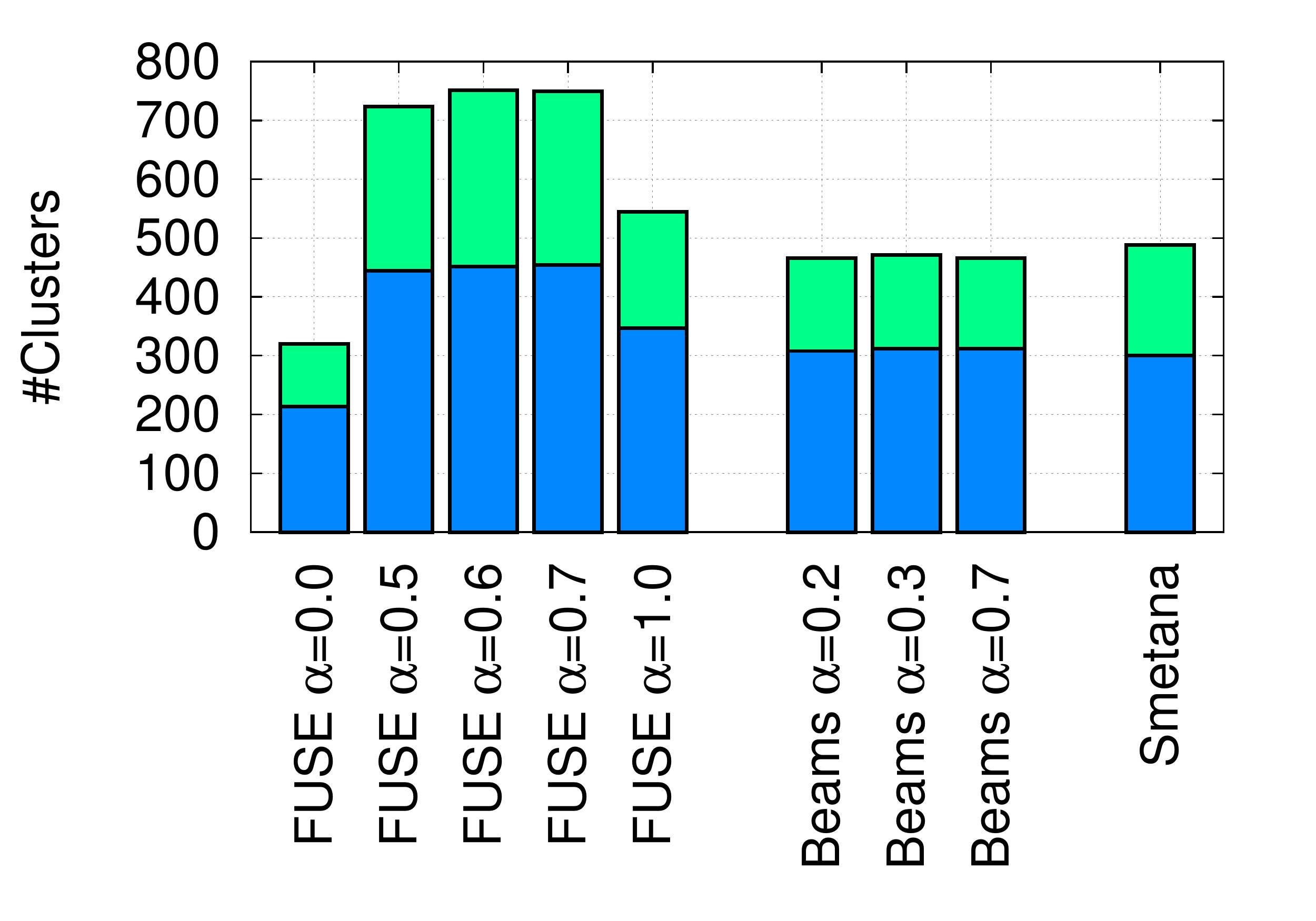} & \hspace{0.5cm}\includegraphics[width=4cm, angle=270]{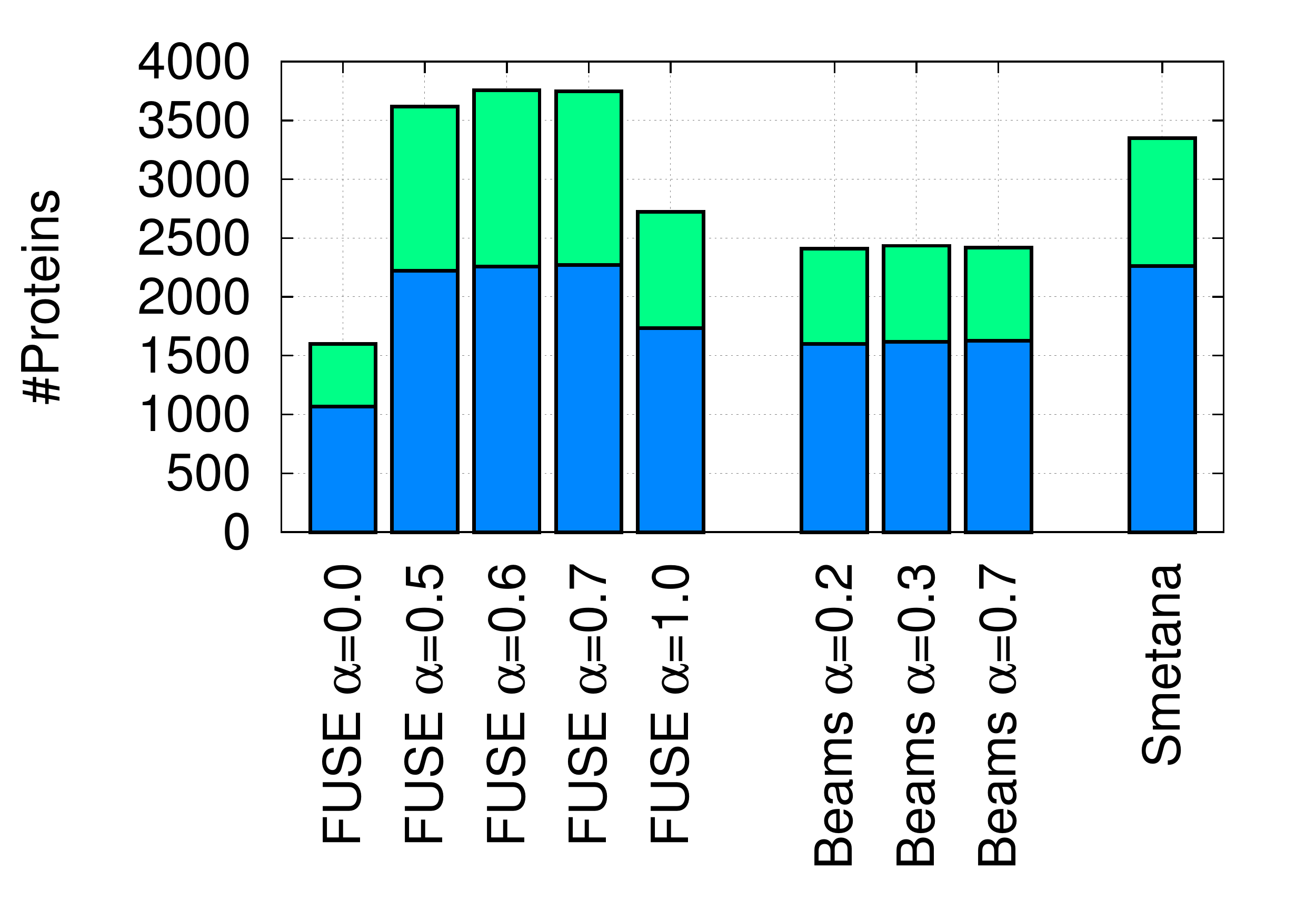} \\
	\end{tabular}
	\caption{{\bf Functional consistency analysis}. Left: for each alignment produced by the compared aligners (for a specific value of $\alpha$ for FUSE and Beams),
	the bar chart shows the number of clusters that contain proteins from all five species and that are BP consistent (in green) or MF consistent (in blue).
	Right: the figure shows the same, but in terms of the number of proteins in these clusters.}
	\label{fig:consistencycomp}
	\end{center}
\end{figure}

\paragraph{\bf Evaluation based on functional consistency.}
We assess functional homogeneity of the clusters obtained by each method.
We say that a cluster is {\em annotated} if at least two of its proteins are annotated by a GO term.
We say that an annotated cluster is {\em consistent} if all of its annotated proteins have at least one common GO term.
The ratio of all consistent clusters to all annotated clusters we call {\em specificity}.
Another consistency measure that is used in previous studies \cite{liao09,sahraeian13,alkan14} is the {\em mean normalized entropy} (MNE).
The normalized entropy of an annotated cluster $c$ is defined as $\displaystyle  NE(c) = - \frac{1}{\log d} ~ \sum^{d}_{i=1} p_i \times \log p_i $,
where $p_i$ is the fraction of proteins in $c$ with the annotation $GO_i$,
and $d$ represents the number of different $GO$ annotations in $c$. MNE is the average of the normalized entropy of all annotated clusters.
We compare FUSE, Beams and Smetana on their ability to uncover functionally conserved proteins across all input networks,
by measuring the consistency, specificity and MNE of their clusters that contain proteins from all five networks (see Figure \ref{fig:consistencycomp} and supplementary material Table 1.
We consider GO annotations from biological process (BP) and molecular function (MF) separately and do not consider cellular component (CC) annotations,
as CC only annotate 9.7\% of the proteins in the five networks.

For BP, FUSE (with $\alpha = 0.6$) outperforms other aligners:
it creates the largest number of consistent clusters, the largest number of proteins in these consistent clusters, the highest specificity, and the lowest MNE.
For MF, FUSE (with $\alpha = 0.7$) obtains the largest number of consistent clusters and of proteins within these consistent clusters, but it is outperformed by Beams in terms of specificity
and MNE. However, Beams achieves these higher specificity and lower MNE by producing smaller number of MF consistent clusters.

Finally, FUSE, obtains the best results when using a combination of sequence similarities and NMTF predicted similarities.
Including predicted similarities ($\alpha=0.6$) allows for finding up to 51\% more of BP consistent clusters and up to 31\% more of MF consistent clusters
than when using sequence similarity alone ($\alpha=1$).
Also, we note that these larger numbers of consistent clusters and proteins are not obtained at the cost of specificity (see supplementary material Table 1.).

FUSE is also computationally efficient and scalable. The matrix factorization step is an $O(\nu^3)$ time operation, where $\nu$ is the number of proteins
in the largest PPI network. On our dataset, the matrix factorization step is the most time consuming, and requires $\approx$ 20 hours 
to complete. The alignment step has a smaller time complexity of $O(kn^2 \log{n} + kne)$, where $n$ is the number nodes in FUSE's $k$-partite graph (i.e., the total number of proteins in all PPI networks),
and $e$ is the total number of edges in FUSE's $k$-partite graph, and on our dataset the alignment process requires less than 15 minutes.
The time complexity of Beams is $O(nd^{k+1})$, where $d$ is the maximum degree of a node in Beams' $k$-partite graph.
Beams complexity becomes larger than FUSE's one when its $k$-partite graph becomes denser (i.e., when $d$ tends to $n$). Aligning our PPI networks with Beams requires $\approx$ 78 hours.
Finally, Smetana has the lowest time complexity of $O(k^3ne)$, and on our dataset it requires $\approx$ 1 hours, but produces an alignment that have lower functional consistency than the ones of FUSE.

\section{Conclusions}

In this paper we propose FUSE, a novel global multiple network alignment
algorithm which can efficiently align even the largest currently available 
PPI networks. FUSE computes novel similarity scores between the proteins in
PPI networks by fusing all PPI network topologies and their protein
sequence similarities by using non-negative matrix tri-factorization. We
show that these new similarities complement solely sequence-based ones: 
NMTF predicts as similar 38 times more protein pairs than sequence alone 
does and these predicted protein pairs are functionally consistent. This 
demonstrates the power of data integration and contribution of network 
topology to sequence-based methods for finding functionally consistent proteins 
in different species.

We define new functional similarity scores between the proteins by combining the
similarity scores obtained by NMTF with the sequence-based ones using a user-defined 
balancing parameter $\alpha$ to either favour one or the other. FUSE uses these functional 
scores to construct global one-to-one multiple network alignment by using a novel maximum 
weight k-partite matching heuristic algorithm.

We compare the alignments of FUSE to the ones of the state-of-the art
aligners, Beams and Smetana, and find that when considering the clusters
that cover all input networks, FUSE produces the largest number of
functionally homogeneous clusters. Additionally, we find that the
results of FUSE are superior to those of the other state-of-the-art aligners when
functional similarity scores are created both from sequence and NMTF scores
(when $\alpha=0.6$) rather than when we use sequence information only (when
$\alpha=1$).  This again demonstrates complementarity of network topology and 
sequence in carrying biological information.\\
\noindent

\section*{Acknowledgments}

\paragraph{Funding} This work was supported by the European Research Council
(ERC) Starting Independent Researcher Grant 278212, the
National Science Foundation (NSF) Cyber-Enabled Discovery
and Innovation (CDI) OIA-1028394, the ARRS project J1-5454, and the Serbian Ministry of
Education and Science Project III44006.

\paragraph{Conflict of interest} None declared.

\bibliographystyle{splncs03}
\bibliography{biblio}

\vspace{7.5cm}
\begin{center}
{\Large \bf Supplementary Material}
\end{center}

\begin{table}[!th]
	\footnotesize
	\begin{center}
	\begin{tabular}{| l l | c c c c c | c c c | c |}
		\hline
		~	& ~& \multicolumn{5}{| c |}{FUSE}								& \multicolumn{3}{| c |}{Beams}				& Smetana\\
			& ~			& $\alpha = $0	& 0.5		& 0.6		& 0.7		& 1		& $\alpha = $0.2	& 0.3		& 0.7		& ~ \\
		\hline
		BP:	& \#C			&  ~ 107 ~~ 	& 280 ~~ 	& \bf 300 ~~ 	& 295 ~~ 	& 198 ~ 	&  ~ 158 ~ 		& 159 ~~ 	& 154 ~ 	&  ~ 188\\
		~	& \#P			&  ~ 535 ~~ 	& 1,400 ~~ 	& \bf 1,500 ~~ 	& 1,475 ~~ 	& 990 ~ 	&  ~ 809 ~ 		& 815 ~~ 	& 790 ~ 	&  ~ 1,086\\
		~	& Spec. 		&  ~ 10.6\% ~~ & 18.6\% ~~ 	& \bf 19.7\% ~~ & 19.5\% ~~ 	& 18.2\% ~ 	&  ~ 19.2\% ~		& 19.4\% ~~ 	& 18.8\% ~ 	&  ~ 14.7\%\\
		~	& MNE			&  ~ 2.38 ~~ 	& 2.15 ~~ 	& 2.14 ~~ 	& 2.14 ~~ 	& \bf 2.11 ~ 	&  ~ 2.19 ~ 		& 2.22 ~~ 	& 2.22 ~ 	&  ~ 2.16\\
		\hline
		MF: 	& \#C			&  ~ 213 ~~ 	& 444 ~~ 	& 451 ~~ 	& \bf 454 ~~ 	& 347 ~ 	&  ~ 308 ~ 		& 312 ~~ 	& 312 ~ 	&  ~ 300\\
		~	& \#P 			&  ~ 1,065 ~~ 	& 2,220 ~~ 	& 2,255 ~~ 	& \bf 2,270 ~~ 	& 1,735	 ~	&  ~ 1,601 ~ 		& 1,619 ~~ 	& 1,628 ~ 	&  ~ 2,262\\
		~	& Spec. 		&  ~ 38.6\% ~~ & 58.3\% ~~ 	& 59.0\% ~~ 	& 59.5\% ~~ 	& 59.9\% ~	&  ~ 65.8\% ~ 		& 66.0\% ~~ 	& \bf 68.3\% ~ 	&  ~ 42.1\%\\
		~	& MNE			&  ~ 0.87 ~~ 	& 0.78 ~~ 	& 0.78 ~~ 	& 0.78 ~~ 	& 0.77 ~ 	&  ~ \bf 0.72 ~ 	& 0.75 ~~ 	& 0.73 ~ 	&  ~ 0.80\\
		\hline
	\end{tabular}
	\vspace{0.1cm}\\
	\parbox{0.99\textwidth}{{\bf Supplementary Table 1. Functional consistency analysis.} Each column represents one of the compared aligners (for a specific value of $\alpha$ for FUSE and Beams).
	Numbers in cell report (from top to bottom): the number of consistent clusters (\#C), the number of proteins in consistent clusters (\#P), the specificity (Spec.), and the mean normalized entropy (MNE).
	In each row, the highest value is shown in bold. }
	\end{center}
\end{table}

\end{document}